\documentclass[journal]{IEEEtran}

\usepackage{xcolor,soul,framed} 

\colorlet{shadecolor}{yellow}
\usepackage[pdftex]{graphicx}
\graphicspath{{../pdf/}{../jpeg/}}
\DeclareGraphicsExtensions{.pdf,.jpeg,.png}

\usepackage[cmex10]{amsmath}
\usepackage{array}
\usepackage{mdwmath}
\usepackage{mdwtab}
\usepackage{eqparbox}
\usepackage{url}
\usepackage{color}
\usepackage{cite}
\usepackage{amssymb,amsfonts}
\usepackage{multirow}
\usepackage{subcaption} 
\usepackage{float} 
\usepackage{algorithm}
\usepackage{algpseudocode}
\usepackage{color}
\usepackage{booktabs}
\usepackage{lineno}
\usepackage{hyperref}
\begin{document}
\bstctlcite{IEEEexample:BSTcontrol}
\title{Physics-Informed Neural Network with Adaptive Clustering Learning Mechanism for Information Popularity Prediction}
\author{
\IEEEcompsocitemizethanks{}}
\author{Guangyin Jin*,~\IEEEmembership{Senior Member,~IEEE}, Xiaohan Ni*, Yanjie Song,~\IEEEmembership{Member,~IEEE}, \\ Kun Wei, Jie Zhao, Leiming Jia, Witold Pedrycz,~\IEEEmembership{Fellow,~IEEE}
\IEEEcompsocitemizethanks{\IEEEcompsocthanksitem This research was supported by the National Natural Science Foundation of China No. 72501042, 72501214, 72601001. The Science and Technology Innovation Program of Hunan Province(2025RC3111).\protect\\
\IEEEcompsocthanksitem  Guangyin Jin, Kun Wei, Jie Zhao and Leiming Jia are with National Innovative Institute of Defense Technology, Academy of Military Science, Beijing 100071, China (E-mail: jinguangyin18@alumni.nudt.edu.cn, weikun@pku.edu.cn, zhaojie12@alumni.nudt.edu.cn, lm\_jia@foxmail.com).
Xiaohan Ni is with Capital Normal University, Beijing, China (E-mail: nixiaohan97@163.com).
Yanjie Song is with the School of Information Science and Technology, Dalian Maritime University, Dalian 116026, China (E-mail: songyj\_2017@163.com).
Witold Pedrycz is with the Silesian University of Technology (SUT), Department of Measurement and Control Systems, Gliwice, Akademicka 2, 44-100 Poland, Department of Electrical and Computer Engineering, University of Alberta, Edmonton, AB T6G 2R3, Canada, Constructor University, Bremen, Germany, also with the Research Center of Performance and Productivity Analysis, Istinye University, Istanbul, Türkiye (E-mail: wpedrycz@ualberta.ca).(Corresponding author: Yanjie Song) \protect\\
\IEEEcompsocthanksitem ${*}$ Both authors contributed equally to this research.
}}


\maketitle

\begin{abstract}
With society entering the Internet era, the volume and speed of data and information have been increasing. Predicting the popularity of information cascades can help with high-value information delivery and public opinion monitoring on the internet platforms. The current state-of-the-art models for predicting information popularity utilize deep learning methods such as graph convolution networks (GCNs) and recurrent neural networks (RNNs) to capture early cascades and temporal features to predict their popularity increments. However, these previous methods mainly focus on the micro features of information cascades, neglecting their general macroscopic patterns. Furthermore, they also lack consideration of the impact of information heterogeneity on spread popularity. To overcome these limitations, we propose a physics-informed neural network with adaptive clustering learning mechanism, PIACN, for predicting the popularity of information cascades. Our proposed model not only models the macroscopic patterns of information dissemination through physics-informed approach for the first time but also considers the influence of information heterogeneity through an adaptive clustering learning mechanism. Extensive experimental results on three real-world datasets demonstrate that our model significantly outperforms other state-of-the-art methods in predicting information popularity.
\end{abstract}


%
\IEEEpeerreviewmaketitle


\section{Introduction}\label{sec:intro}
\IEEEPARstart{W}{e} are in an era of information explosion. According to the statistics\footnote{https://www.statista.com}, from 2010 to 2023, the amount of information generated wordwide every day has increased by 60 times, from about 2 zettabytes to about 120 zettabytes. Therefore, screening high-value or potentially disseminating information plays an important role in online media platform information recommendation and public opinion pre-warning applications. The core of these applications is to accurately predict the popularity of information cascades. Predicting the popularity of information cascades is a fundamental yet challenging task.{Over the past extended period, various methods based on mathematical modeling with social physics~\cite{jusup2022social}, and machine learning with feature engineering~\cite{zhou2021survey} have been successively applied in this field. However, due to the highly nonlinear nature and noise impact during the information propagation process in complex networks~\cite{ji2023signal}, combined with complex network characteristics such as small-world and power-law properties~\cite{artime2024robustness}, these traditional methods struggle to achieve breakthroughs in accuracy.}
In recent years, more and more deep learning models have been proposed for predicting information popularity~\cite{gao2020deep}. In particular, some recent state-of-the-art methods~\cite{liao2019popularity,tai2023predicting,bao2024popularity,zhong2023hierarchical,chen2019information} capture the topological features of information cascades using models from graph convolutional networks (GCNs) series, and combine them with sequence modeling methods such as recurrent neural networks (RNNs) to capture the dynamics within observable time intervals. Although these deep learning models have been demonstrated effective in information popularity prediction, there are still at least two limitations.

\begin{figure}[t]
\centering
\vspace{-3mm}
\includegraphics[width=0.48 \textwidth]{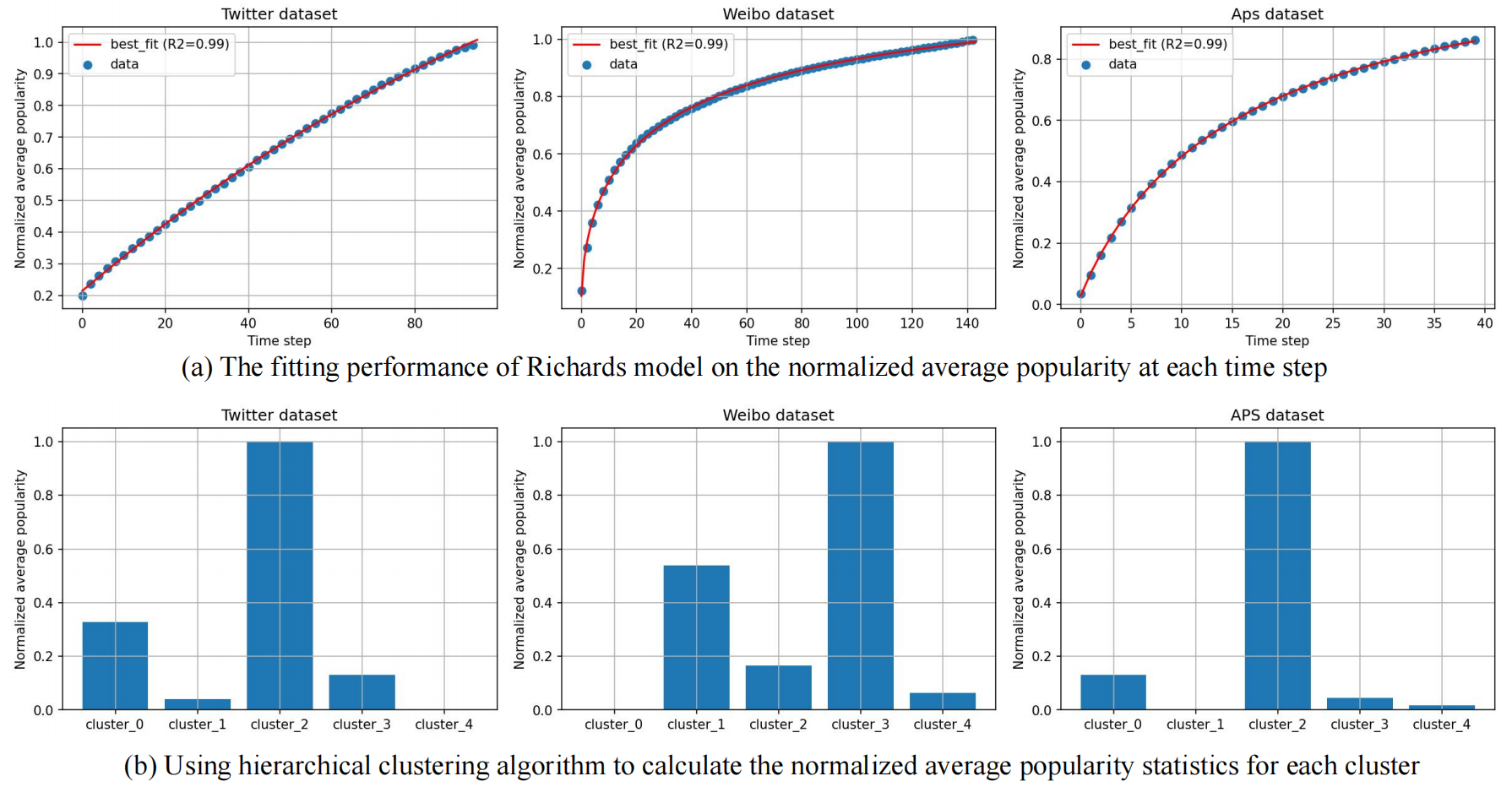}
\caption{Nonlinear function fitting and clustering performance on Twitter, Weibo and APS dataset.}
\label{fig:intro} 
\vspace{-5mm}
\end{figure}

(a) \textbf{Lack of modeling of universal physical patterns.} The most advanced deep learning models recently used for predicting information popularity have focused more on how to better learn micro-level features, such as cascade patterns and temporal patterns within observable intervals. However, they have overlooked the macroscopic physical laws, such as the temporal dynamics of information propagation, whose physical characteristics may be one of the most crucial factors determining the upper limit of information popularity. Based on some previous works~\cite{wang2012diffusive,sarkar2017understanding,jin2025will}, the spread of information is similar to the growth process of a biological organism, where the spread initially increases rapidly, then gradually levels off and reaches an upper limit. The curve of information spread resembles an irregular S-shape. Many studies on rumor dynamics have also utilized classic epidemic models such as SI and SIR to model the spread of rumors~\cite{moreno2004dynamics,yu2021modeling,wang2019global}. In these models, the growth pattern of rumor spreaders also approximates an S-curve. As shown in Fig.~\ref{fig:intro}, the blue data points represent the average popularity of information at each time step in different datasets. We fit these data points using the flexible S-shaped curve model called Richards model~\cite{farthing2017numerical}. The Richards model effectively fits the data points in the Weibo, Twitter, and APS datasets, as shown by the solid red line. The R-squared values for all datasets reached 0.99. This indicates that information propagation at the macroscopic level conforms to the physical laws represented by the Richards function.

(b) \textbf{Lack of consideration of the impact of information heterogeneity.} The patterns of information propagation exhibit both universality and diversity. While the macroscopic patterns of information propagation may follow variants of the S-shaped curve, the influence of information heterogeneity on its popularity cannot be ignored. According to previous studies~\cite{kim2013modeling,si2020comparative}, different categories of news information, such as political news, entertainment news, and lifestyle information, exhibit significant differences in their propagation patterns. The dissemination of scientific information follows a similar trend. For example, in the APS dataset, it has been found that scientific literature of different categories may display significantly different propagation life-cycles~\cite{zhou2021survey}. As shown in Fig~\ref{fig:intro}, we employ hierarchical clustering algorithm to divide the data into five clusters on three different datasets. We found that the average propagation popularity varies significantly among different clusters, indicating the significant impact of information category heterogeneity.
In recent deep learning research related to information popularity prediction, exploring and uncovering the impact of these factors is still an ongoing endeavor. 

To address the above problems, we propose a novel framework physics-informed neural network with adaptive clustering learning mechanism (PIACN) for information popularity prediction. To be specific, we first utilize a self attention-based cascade embedding network for learning the microscopic cascade dynamics and lightweight CNN-based temporal learning network to capture the macroscopic temporal dynamics of information propagation within observable intervals. Then we design a physical modeling network to approximate the adjustable parameters of the Richards function, aiming to model macroscopic physical laws and constrain the training loss function. Furthermore, we propose incorporating adaptive deep clustering mechanism in the model, where cluster centers are adopted as crucial features supporting the final predictions, taking into account the impact of information heterogeneity. Our main contributions in this paper are summarized as follows:
\begin{itemize}
\item We first propose to employ the Richards function to model the physical laws of information propagation at a macroscopic level and design a physics-informed network to nonlinearly approximate the learnable parameters of Richards function. During the model training process, we embedded physical constraints into the loss function to guide the neural network towards learning in a direction that aligns with the objective laws. 
\item We first propose to involve the adaptive clustering learning mechanism into the information popularity prediction model, fully considering the impact of information category heterogeneity on the prediction results.
\item We conduct extensive experiments on three public datasets. The experimental results demonstrates that our model can obtain more than 12\% improvements compared with the state-of-art baselines.
\end{itemize}

The rest of this paper is organized as follows. We first review the related works about information popularity prediction and physical-informed neural networks in Section~\ref{sec:related_work}. Then the definition of information popularity prediction is formulated in Section~\ref{sec:define}. Motivated by the presented challenges, we introduce the details of our methodology in Section~\ref{sec:method}. After that, we design sufficient experiments to evaluate our model from multiple perspectives in Section~\ref{sec:exp}, where the ablation studies, case studies, efficiency analysis and parameter studies are conducted. Finally, we conclude our contributions and future directions in Section~\ref{sec:conclusion}.

\section{Related Work}\label{sec:related_work}
\subsection{Information Popularity Prediction}
According to the line of existing works, the methods in this field can be roughly divided into three categories: mathematical modeling methods, feature-based machine learning methods, and deep learning methods.

\textbf{Mathematical modeling methods:} The series of mathematical modeling methods adopted dynamics equations to simulate the process of information diffusion. For instances, Zhao et al.~\cite{zhao2015seismic} proposed a self-exciting point process to model the dynamic evolution of information popularity, Shen et al.~\cite{shen2014modeling} utilized a self-enhancing Poisson process to capture forwarding behavior in social networks. Rizoiu et al.~\cite{rizoiu2017expecting} explored the correlation between internal and external information dissemination dynamics by combining the self-exciting point process with the SIR epidemic model. 
The advantage of these methods is that they can establish explicit dynamic models, but they struggle to effectively learn the entanglement and highly nonlinear characteristics information propagation.

\textbf{Feature-based machine learning methods:} 
With the development of gradient boosting tree approaches~\cite{natekin2013gradient}, machine learning models can better equipped to automatically capture the influencing factors of information popularity from diverse sources of heterogeneous features. Huang et al.~\cite{huang2018random} extracted features by combining both metadata of the posts and users and adopted random forest model to effectively predict the social media popularity. Tavazoee et al.~\cite{tavazoee2020recurrent} proposed a recurrent random forest for the assessment of popularity based on the characteristics of content or sentiment information extracted from the tweets. Wang et al.~\cite{wang2020feature} address the issue of missing features in the popularity prediction by utilizing two CatBoost trainers. 
In general, feature-based machine learning methods can achieve a data-driven paradigm without relying on dynamic equations, but they are limited by cumbersome feature engineering.

\textbf{Deep learning methods:} In recent years, deep learning has emerged as a mainstream approach for predicting information popularity. 
DeepCas~\cite{li2017deepcas} is the pioneering deep learning model for information popularity prediction, which leveraged a random walk algorithm to transform the cascade graph into multiple forwarding paths. DeepHawkes~\cite{cao2017deephawkes} achieved a balance between prediction performance and interpretability by utilizing deep learning techniques to capture three crucial factors of the self-exciting point process. DFTC~\cite{liao2019popularity} combines content features of online articles on WeChat with the long-short term dynamics of the diffusion process for popularity prediction. CasCN~\cite{chen2019information} introduced a novel approach by treating the cascade graph as a sequence of sub-graphs and employed a recursive GCN to encode the structural representation in each snapshot. CoupledGNN~\cite{cao2020popularity} proposed to use two coupled graph neural networks to capture the interaction between node activation states and social relationships. 
CasFlow~\cite{xu2021casflow} introduced a hierarchical cascade normalizing flow approach to capture both the cascade representation uncertainty and node infection uncertainty.
TEDDY~\cite{bao2024popularity} enabled GCN to collaborate with multiple sequences temporal encoder and gated recurrent networks to capture multi-scale temporal dynamics of information cascades. Similarly, {PiGCN~\cite{yu2024information} adopted GCN to collaborate with a dynamic neural network for information cascade size prediction.}
CasHAN~\cite{zhong2023hierarchical} introduced a neural network framework with hierarchical attention mechanisms to integrate node-level attention based on user influence and sequence-level attention based on community redundancy.
I3T~\cite{tai2023predicting} partition the information cascade into sequences using DeepWalk and refine the node embeddings concurrently with both GCN and DeepWalk, capturing both inter-path and intra-path influence transitivity.
Although deep learning methods have made significant breakthroughs in predicting information popularity, they still overlook the macroscopic physical laws and heterogeneity of information categories, thus hindering further improvement in predictive performance.

\subsection{Physics-informed Neural Networks}
In recent years, physics-informed neural networks (PINN) has been increasingly applied in the intersection of artificial intelligence, natural sciences, and social sciences~\cite{cuomo2022scientific}. The aim is to combine the nonlinear computational capabilities of neural networks with the fundamental laws that can be mathematically represented in related scientific fields to achieve superior and more interpretable predictive results. Currently, PINN have made significant breakthroughs in fields such as transportation, epidemiology, fluid mechanics and some other dynamical systems. In transportation field, traffic flow differential equations~\cite{shi2021physics,zhang2024physics,jin2025m3} and energy-based differential equations~\cite{ji2022stden} were embedded into deep learning models to guide training phase for traffic state estimation. In addition, gravity models and Hawkes processes are respectively applied in the physics-informed neural network (PINN) models for OD flow prediction~\cite{rong2023origin,simini2021deep} and traffic congestion event prediction~\cite{jin2023spatio,zhang2024deep,zhu2021spatio,jin2025exploring}. In epidemiology, Kapoor et al.~\cite{kapoor2020examining} and Gao et al.~\cite{gao2021stan} combined SIR equation with neural networks for worldwide Covid-19 forecasting. Xie et al.~\cite{xie2022epignn} embedded the meta-population model into a spatio-temporal graph neural network model to predict the spread of epidemics in different spatial locations. In fluid mechanics field, many PINN-based models embed classical equations such as Navier-Stokes and Cahn-Hilliard into neural networks to simulate or complete incompressible high-speed flows or turbulent flows~\cite{yang2022learning,cai2021physics,eivazi2024physics,Ren2026Evolutionary}. 
In some other dynamic systems, such as power systems~\cite{misyris2021capturing}, Gray-Scott systems~\cite{giampaolo2022physics} and industrial control systems~\cite{arnold2021state}, PINNs have demonstrated their advantages by outperforming simple dynamic models and pure neural network methods in prediction or simulation.

\section {Preliminary}\label{sec:define}

\begin{figure}[t]
\centering
\vspace{-3mm}
\includegraphics[width=0.46 \textwidth]{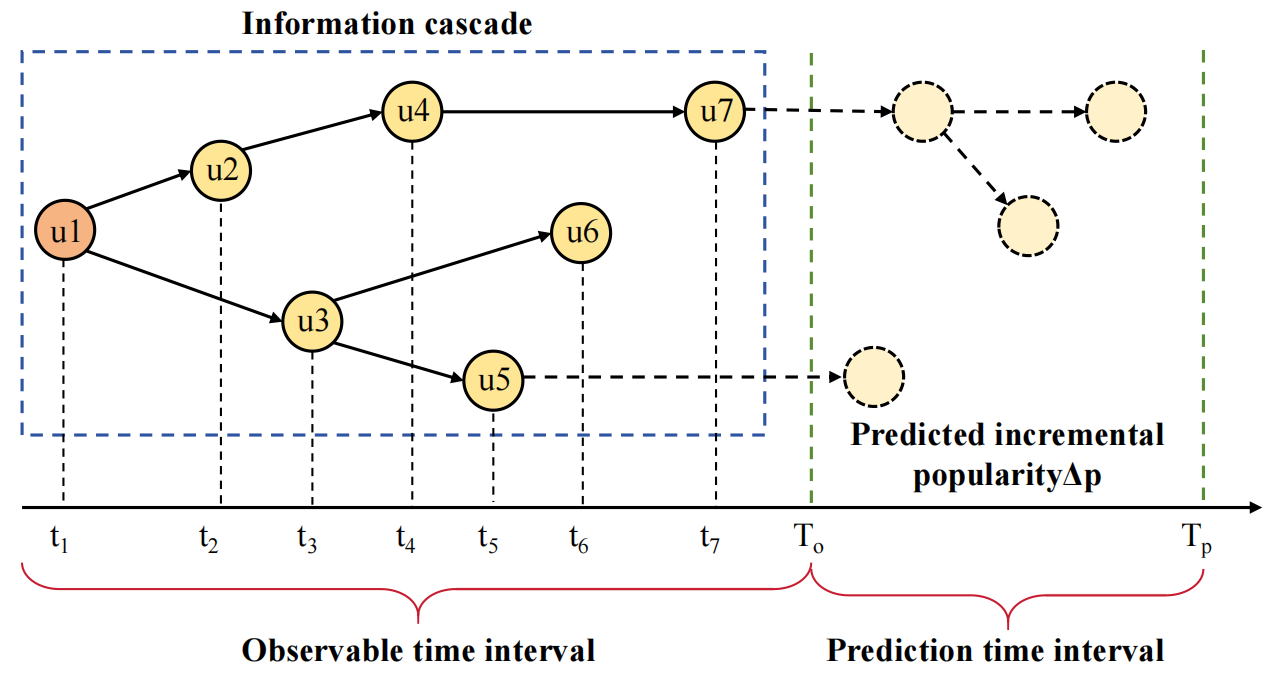}
\caption{The formulation of information popularity prediction.}
\label{fig:task} 
\vspace{-3mm}
\end{figure}

\subsection{Problem Definition}
The description of the information cascade popularity prediction task in this paper is illustrated in Fig.~\ref{fig:task}. As shown in this figure, a message is initially posted by the original user $u_{1}$ and then forwarded by other users. For each message, an observable time interval $T_{o}$ and a prediction time interval $T_{p}$ are set. We utilize the historical cascading features of information propagation occurring before $T_{o}$ to train the prediction model. Let $P_{o}$ and $P_{p}$ represent the information popularity at moments $T_{o}$ and $T_{p}$ respectively. The main task of this study is to predict the incremental popularity $\Delta P = P_{p} - P_{o}$. 
It should be noted that in most previous works, due to concerns about significant differences in data scales, it is common practice to logarithmically transform both the predicted results and the labels.

In most deep learning frameworks, to facilitate data retrieval, it is common practice to divide continuous time intervals into multiple non-overlapping time snapshots. Simultaneously, various features of information cascades evolving with time snapshots need to be considered, such as user characteristics $u_t$, propagation cascade structures $G_t$, and current popularity $p_t$ in time snapshot $t$. Through this series of input features and the deep learning predictor $f(\cdot)$ parameterized by $\theta$, future popularity is predicted:
\begin{equation}
f\left(\left\langle u_1, p_1, G_1\right\rangle, \cdots,\left\langle u_t, p_t, G_t\right\rangle, \theta \mid t<T_o\right) \rightarrow \Delta P
\end{equation}

\subsection{Richards Growth Equation}
\begin{figure}[t]
\centering
\vspace{-3mm}
\includegraphics[width=0.45 \textwidth]{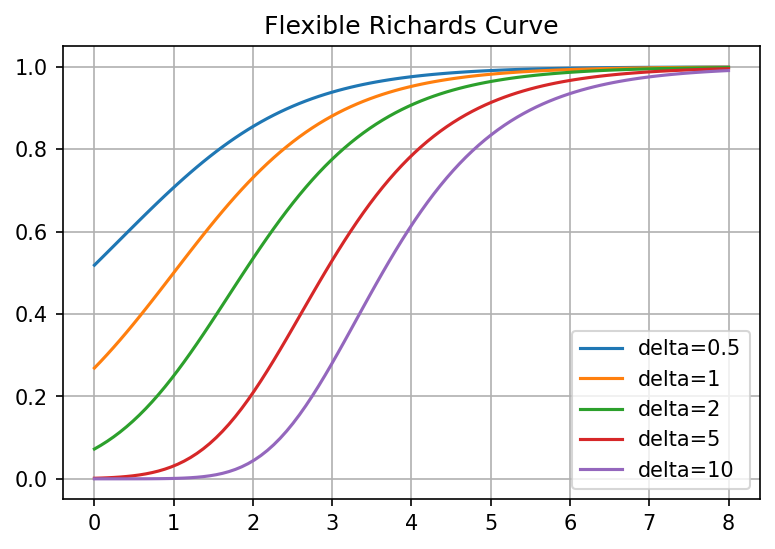}
\caption{The shape of the Richards growth curve changes with the variation of the parameter $\delta$.}
\label{fig:Richards} 
\vspace{-3mm}
\end{figure}

Richards growth equation is a mathematical model originally used in hydrology and soil science to describe the movement of water in unsaturated soils~\cite{weill2009generalized}. This model provides a non-linear equation that captures the dynamics of water flow through soil under varying moisture conditions, which balances the fluxes of water due to gravity, capillarity, and soil water content gradients within the soil profile. 
Richards growth equation can be seen as a member of the logistic model family, but can be used to model asymmetric S-shaped curves in a more flexible manner, with the following formula:  
\begin{equation}
Y=\alpha /\left[1+e^{\beta-\gamma \cdot x}\right]^{\frac{1}{\delta}}
\label{eq:richards}
\end{equation}
{where $\alpha$ denotes the upper asymptote, $\beta$ denotes the growth rate, $\gamma$ and $\delta$ respectively represent the inflection point control parameter and steepness control parameter.} Assuming $\alpha=1$, $\beta=1$, $\gamma=1$, and only adjusting the value of $\delta$, the shape of Richards curve changes as shown in the Fig.~\ref{fig:Richards}.

\begin{figure*}[h]
\centering
\vspace{-2mm}
\includegraphics[width=0.9 \textwidth]{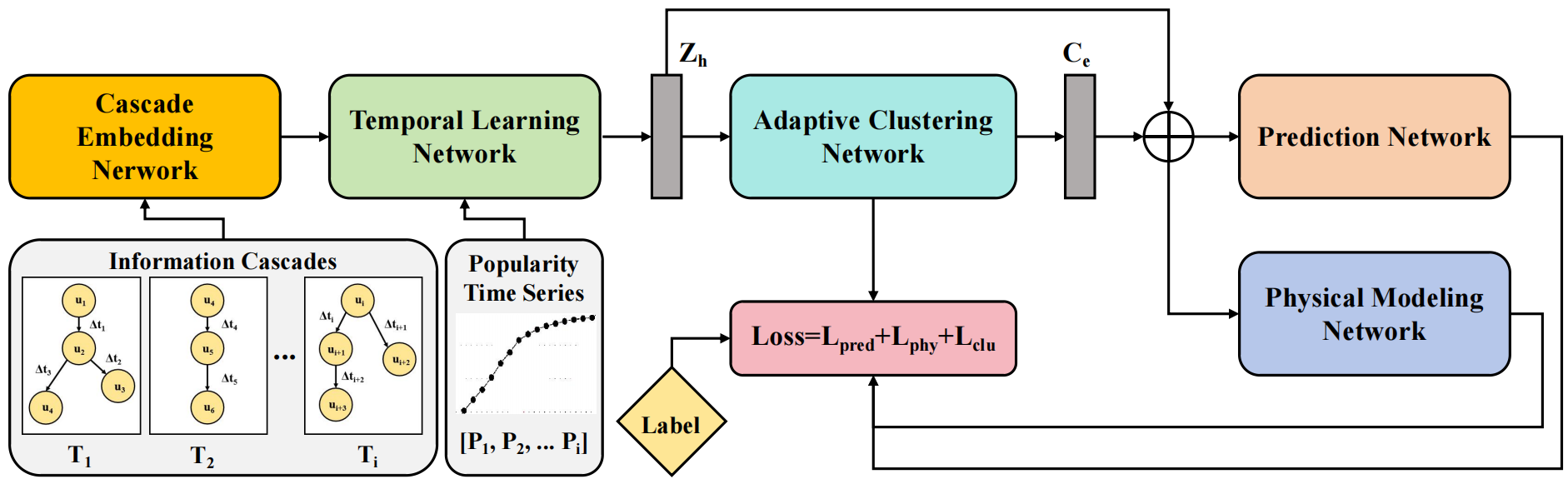}
\caption{The overview of our proposed deep learning model PIACN. {PIACN is composed of cascade embedding network, temporal learning network, adaptive clustering network, prediction network and physical modeling network. There are two parts of input, information cascades and popularity time series, which are put forward into cascade embedding network and temporal learning network respectively. The loss function of our proposed model consists of three parts: prediction loss $L_{pred}$, physical constraint loss $L_{phy}$, and clustering loss $L_{clu}$.}}
\label{fig:overview} 
\vspace{-2mm}
\end{figure*}

\section{Methodology}\label{sec:method}
The overview of our proposed model PIACN is illustrated in Fig.~\ref{fig:overview}. The proposed model consists primarily of five modules: cascade embedding network, temporal learning network, adaptive clustering network, prediction network, and physical network. Cascade embedding network is the first module, which takes as input the features of information cascades from multiple non-overlapping time snapshots, aiming to learn latent representations within the micro-level information cascade in each time snapshot. Temporal learning network is designed to capture the macroscopic evolutionary dynamics of information popularity across different time snapshots. It takes as input the popularity time series and the information cascade representations output by cascade embedding network. The multi-scale hidden representations of each piece of information obtained through cascade embedding network and temporal learning network serve as the input to the adaptive clustering network. This module adaptively clusters the obtained hidden representations and outputs their clustering center representations as distinctive features for discerning the heterogeneity of information categories. The clustering center representations output by adaptive clustering network are summed with the corresponding hidden representations and passed to prediction network and physical modeling network, which respectively output neural predictions and physical predictions. The physical modeling network can automatically learn the parameters of physical equations that represent macroscopic laws and guide the training of the model through the physical boundaries. The loss function of our model comprises three parts: prediction loss, physical constraint loss, and clustering loss. 

\begin{figure}[h]
\centering
 \vspace{-2mm}
\includegraphics[width=0.48 \textwidth]{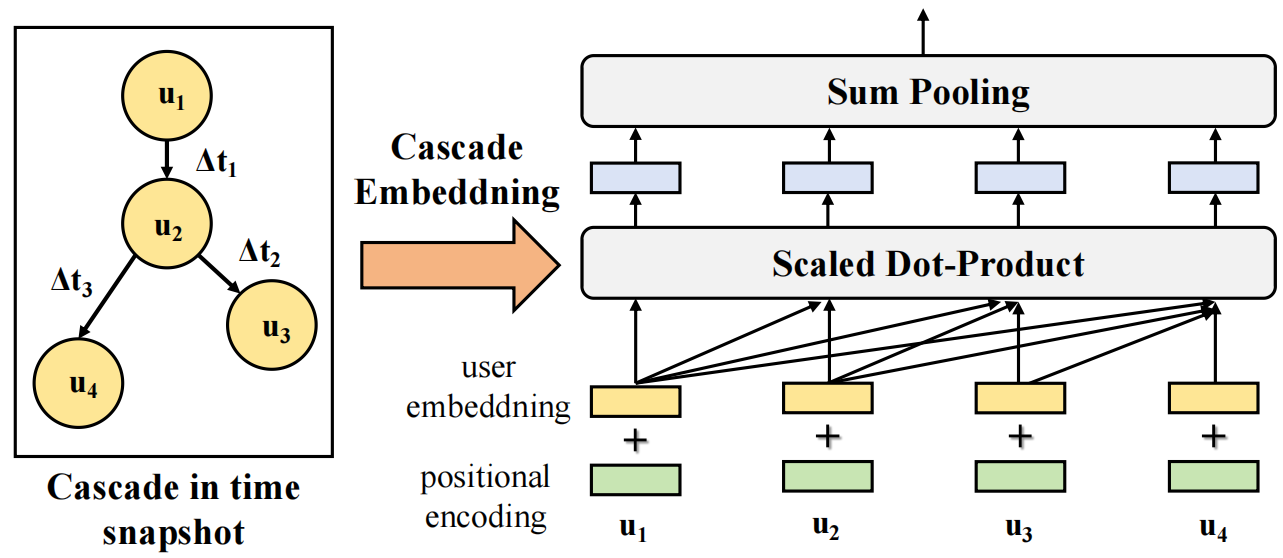}
\caption{The architecture of cascade embedding network. {The dynamics of information cascades are embedded into the latent space by self-attention architecture.}}
\label{fig:cen} 
\vspace{-2mm}
\end{figure}

\subsection{Cascade Embedding Network}
According to many previous works~\cite{van2011opinion, song2007identifying, flodgren2019local}, opinion leaders often play a crucial role in information dissemination in social networks. Information disseminated by them is more likely to be retweeted by a larger number of ordinary users. Additionally, information that is rapidly retweeted in the early stages of dissemination is more likely to stand out and become potentially highly popular information. Therefore, factors such as who spreads the information and how quickly it is retweeted in the early stages significantly influence its future popularity. 
Cascade embedding network aims to embed the important characteristics of information cascades (eg., user features and propagation speed) of each time snapshot into the high-dimensional latent space, enhancing the capability of representation learning. We first embed the micro-temporal features of the propagation cascade as follows:
\begin{equation}
\phi(\Delta t)[s]= \begin{cases}\Delta t, & \text { if } s=0 \\ \sin \left(\omega_k \Delta t\right), & \text { if } s=2 k+1 \\ \cos \left(\omega_k \Delta t\right), & \text { if } s=2 k+2\end{cases}
\end{equation}
where $\Delta t$ represents the time difference in the propagation of information from the previous user to the next user, $\omega_k$ is the learnable parameter for each time difference. Similar to the time positional encoding in Transformer~\cite{vaswani2017attention}, we employ sine and cosine functions to differentiate odd and even positions.
The continuity and monotonicity of the sine and cosine functions ensure that for any two adjacent positions, their corresponding encoding vectors have only minor changes in each dimension, while for any two distant positions, their corresponding encoding vectors exhibit significant differences in each dimension.


Based on the temporal embedding operation, we adopt self-attention architecture to embed the cascades of information propagation, which is formulated as follows:
\begin{align}
Q &= W_q\cdot[u_{e}, \phi(\Delta t)],\\
K &= W_k\cdot[u_{e}, \phi(\Delta t)],\\ 
V &= W_v\cdot[u_{e}, \phi(\Delta t)],\\
Att(Q, K, V)& = Softmax(\frac{Q\cdot K^T}{\sqrt{d}})\cdot V
\end{align}
where $Q$, $K$, $V$ respectively denote the query, key and value in self-attention, $W_q$, $W_k$ and $W_v$ are corresponding learnable weights, $u_{e}$ denotes the embedding of users in information cascades, $d$ is the dimension of self-attention network. In this case, the user embedding $u_{e}$ can be initialized by random vectors if no specific features are provided.
To stabilize the training process of self-attention network, we can also employ the multi-head attention strategy as follows:
\begin{equation}
MultiAtt(Q, K, V) = \Vert_{i=1}^{h} [Softmax(\frac{(Q\cdot K^T}{\sqrt{d|h}})\cdot V]
\end{equation}
where $h$ denotes the number of attention heads and $\Vert$ denotes the concentration operation. 

The architecture of cascade embedding network is shown as Fig.~\ref{fig:cen}. 
The output of the network is to put forward the hidden representation corresponding to each user through a sum pooling operation, and finally obtain the cascade embedding of each time snapshot.

\subsection{Temporal Learning Network}
To obtain the multi-scale hidden representation of information cascades, We should not only focus on the micro-level cascading dynamics within each time snapshot, but also pay attention to the long-term temporal dependencies of information propagation. Considering the high resource consumption and low computational efficiency of RNN-based model and Transformer-based model in capturing long sequence dependencies, we employ a CNN-based architecture in this case for long-term temporal learning. 

\begin{figure}[h]
\centering
\vspace{-2mm}
\includegraphics[width=0.48 \textwidth]{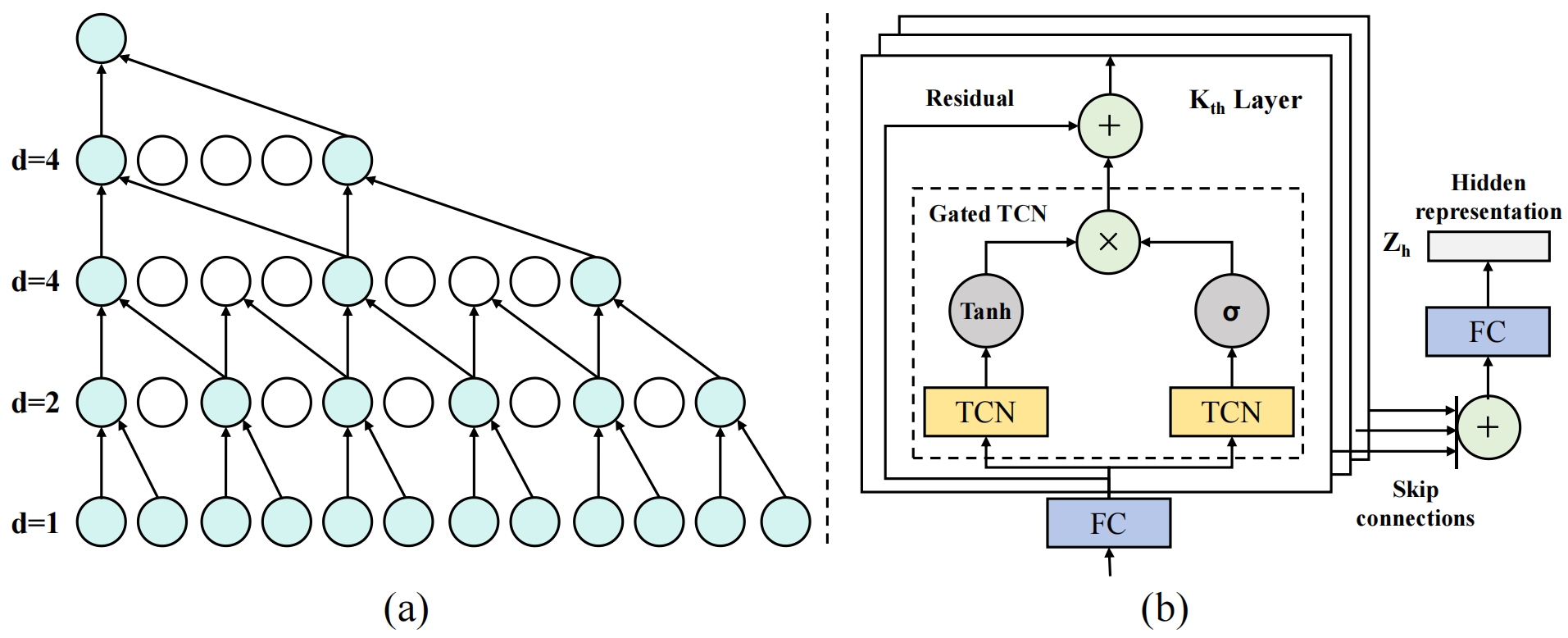}
\caption{The architecture of temporal learning network. {This network consists of multiple layers of gated TCN, residual connections, and skip connections, achieving the learning of long-term temporal dependencies through layer-wise increasing dilation factors.}}
\label{fig:wavenet} 
\vspace{-2mm}
\end{figure}

As illustrated in Fig.~\ref{fig:wavenet}, we utilize a multi-layer dilated temporal convolution architecture for capturing the long-range temporal dynamics. The diagram of multi-layer dilated convolution is shown in Fig.~\ref{fig:wavenet}(a), where each circle represents an input at an independent time step, and white circles represent the dilation intervals, with the dilation interval equal to the value of the dilation factor minus one. By involving the dilation factor, the receptive field of the convolution increases, allowing the model to shift its focus from short-range dependencies among neighboring time steps to long-range dependencies across larger intervals.
Similar to~\cite{yu2016multi}, we can obtain the larger receptive field by expanding the dilation factor in temporal convolution when the layer goes deeper. In this manner, the temporal learning network can capture the short and long-range dependencies in information cascades layer-by-layer.
The dilated temporal convolution operation is defined as:
\begin{equation}
    x \star \mathbf{f}(t) = \sum_{s=0}^{k-1} \mathbf{f}(\theta)\cdot x(t-d\times s),
\end{equation}
where $x\in \mathbf{R}^T$ denotes the given 1D sequence input, $\mathbf{f} \in \mathbf{R}^K$ denotes a temporal convolutional filter at step $t$, $\theta$ denotes the learnable weights of the filter and $d$ denotes the dilation factor. 

The specific architecture of the temporal learning network is shown in Fig.~\ref{fig:wavenet}(b), where each layer consists of a gated temporal convolutional network and a residual connection. The gating mechanism controls the flow of information put forward to the next layer, while the residual connection preserves shallow features of the neural network. By stacking the same neural architecture for $k$ layers with an increasing dilation factor per layer, the temporal learning network have the capability to effectively capture short-term and long-term information cascades. The multi-layer network ultimately outputs its learned hidden representations through skip connections and a fully connected layer. The computational process of the entire network is as follows:
\begin{align}
z_0 &= W_i\cdot[z_t, p_t] + b_i,\\
z_k &= \mathbf{tanh}(\mathbf{\theta_k^1}\star z_{k-1})\odot\mathbf{\sigma}(\mathbf{\theta_k^2}\star z_{k-1}) + z_{k-1}, \\  
Z_h &= W_o\cdot (\sum\nolimits_{k} W_k\cdot z_k) + b_o
\end{align}
where $z_t$ and $p_t$ are respectively the hidden representation from cascade embedding network and the popularity time series, $W_i$ and $b_i$ are the parameters of the input fully connected layer, $z_0$ and $z_k$ are respectively the initial input of the gated temporal convolutional network and the output of the $k_{th}$ layer. 
$\mathbf{\theta_k^1}$ and $\mathbf{\theta_k^2}$ denote the parameters of convolutional kernel, $\odot$ is the element-wise product, $tanh(\cdot)$ is the tanh activation function of the outputs, and $\sigma(\cdot)$ is the sigmoid function which controls the ratio of information flow put forward to the next layer. $W_k$ is the weight parameter of the skip connection of the $k_{th}$ layer, $W_o$ and $b_o$ are the parameters of the output fully connected layer, and $Z_h$ denotes the hidden representation output from the temporal learning network.

\subsection{Adaptive Clustering Network}
Due to the heterogeneity of information categories, which may lead to significant differences in information propagation patterns, we attempt to use information category as an important feature to facilitate model learning. However, lacking labels for information categories, we can only perform clustering learning on them in an unsupervised manner. Therefore, we need to develop an adaptive clustering method that surpasses classic algorithms like K-means, allowing for integration into end-to-end deep learning frameworks. We propose to use the Student’s t-distribution~\cite{van2008visualizing} as a kernel to measure the similarity between the data representation and the cluster center vector.  For the $i_{th}$ sample and $j_{th}$ cluster, the similarity measurement is formulated as follows:
\begin{equation}
q_{i j}=\frac{\left(1+\left\|\mathbf{z}_i-\boldsymbol{\mu}_j\right\|^2 / v\right)^{-\frac{v+1}{2}}}{\sum_{j^{\prime}}\left(1+\left\|\mathbf{z}_i-\boldsymbol{\mu}_{j^{\prime}}\right\|^2 / v\right)^{-\frac{v+1}{2}}},
\end{equation}
where $\mathbf{z}_i$ is the hidden representation of an information sample, $\boldsymbol{\mu}_j$ is the adaptive cluster center vector that can be learned during the neural back-propagation process, $\sum_{j^{\prime}}$ denotes the the collective set of all cluster center vectors. $v$ is the degree of freedom of the Student's t-distribution and $q_{ij}$ can be treated as the probability of assigning data representation sample $i$ to cluster $j$.

After obtaining the cluster probability distribution $q_{i j}$, our goal is to optimize data representation by learning from high-confidence assignments. To be specific, we aim to push the data representation closer to the cluster centers, thereby enhancing clustering cohesion. Therefore, we compute the target distribution $p_{ij}$ as follows:
\begin{equation}
p_{ij}=\frac{q_{i j}^2 / \sum_i q_{ij}}{\sum_{j^{\prime}} q_{i j^{\prime}}^2 / \sum_i q_{ij^{\prime}}},
\end{equation}
where $\sum_i q_{ij^{\prime}}$ denotes the soft cluster frequencies. In the target distribution, each assignment $q_{ij}$ is squared and then normalized to increase the confidence of the assignments. To make the student t-distribution more approximate to the target distribution, the objective function is formulated using the Kullback-Leibler (KL) divergence and defined as follows:
\begin{equation}
\mathcal{L}_{clu}=\sum_i \sum_j p_{ij} \log \frac{p_{i j}}{q_{i j}}
\label{eq:clu_loss}
\end{equation}
By minimizing the KL divergence loss between Student's t-distribution and the target distribution, the neural network can adaptively optimize the cluster center vector $\boldsymbol{\mu}_j$, thus thus achieving end-to-end data representation clustering. 
As shown in Fig.~\ref{fig:overview}, the clustering center vectors output from adaptive clustering network are added to the corresponding class-specific hidden representations output from temporal learning network, enhancing the original hidden representations. The calculation process is as follows:
\begin{equation}
Z_p = Z_{h}\bigoplus C_{e}
\end{equation}
where $C_{e}$ and $Z_{h}$ are respectively the clustering center tensor and the hidden representations output from temporal learning network, $\bigoplus$ means the class-specific sum, $Z_p$ is the output representation that will be fed into prediction network and physical modeling network. 

\subsection{Prediction \& Physical Modeling Network }
In order to incorporate macroscopic physical constraints into the deep learning model for prediction, we propose to design the prediction network and physics modeling network to respectively output neural predictions and physical predictions. The prediction network is a simple two-layer fully connected network, which directly output the predictions based on the hidden representation $Z_p$, which is expressed as:
\begin{equation}
Y_{pred} = W_{p2}\cdot(ReLU(W_{p1}\cdot Z_p + b_{p1}))+b_{p2},
\end{equation}
The physics modeling network approximates the learnable parameters in Eq.~\ref{eq:richards} Richards growth equation through multiple fully connected neural networks, enabling adaptive learning of physical parameters through the back-propagation training process of neural networks. Since the physical parameters in this equation need to be strictly positive to ensure physical significance, we use Softplus function to transform the output layer of the results, which are formulated as follows:
\begin{align}
\alpha &= \mathbf{ln}(1+\mathbf{exp}(W_{\alpha}\cdot Z_p + b_{\alpha})),\\
\beta &= \mathbf{ln}(1+\mathbf{exp}(W_{\beta}\cdot Z_p + b_{\beta})),\\
\gamma &= \mathbf{ln}(1+\mathbf{exp}(W_{\gamma}\cdot Z_p + b_{\gamma})),\\
\delta &= \mathbf{ln}(1+\mathbf{exp}(W_{\delta}\cdot Z_p + b_{\delta})),
\end{align}
{where $\alpha$, $\beta$, $\gamma$ and $\delta$ are the four important parameters in the Richards equation, representing upper asymptote, growth rate, inflection point control parameter, and steepness control parameter respectively.}
$W_{\alpha}$, $b_{\alpha}$, $W_{\beta}$, $b_{\beta}$, $W_{\gamma}$, $b_{\gamma}$, $W_{\delta}$ and $b_{\delta}$ are weights and bias of fully connected neural networks in the physics modeling network. Based on these learned parameters, the incremental popularity can be computed by the physics modeling network as follows:
\begin{equation}
Y_{phy} = \mathbf{ln}(\alpha /\left[1+\mathbf{exp}({\beta-\gamma \cdot t})\right]^{\frac{1}{\delta}}- Y_{T})    
\end{equation}
where $t$ denotes the time steps. Based on the learned Richards growth equation, the corresponding incremental popularity $Y_{phy}$ can be output in a generative manner by providing the corresponding time step index.

\subsection{Loss Function}
In our model, the loss function consists of three components: prediction loss, physical constraint loss, and clustering loss. The prediction loss is the mean absolute error between the ground truths and prediction results, which is as follows:
\begin{equation}
\mathcal{L}_{pred} = \frac{1}{N}\sum_{i=0}^{N}\|Y^{i} - Y_{pred}^{i}\|,
\end{equation}
where $N$ denotes the number of samples, $Y^{i}$ and $Y_{pred}^{i}$ are respectively the real incremental popularity and the neural prediction.

The physical constraint loss can be further divided into two parts: reconstruction constraint part and prediction constraint part, which is formulated as:
\begin{align}
\mathcal{L}_{rc} &= \frac{1}{N}\frac{1}{T}\sum_{i}^{N}\sum_{t}^{T}\|Y_{t}^{i} - \hat{Y}_{t}^{i}\|,\\
\mathcal{L}_{pc} &= \frac{1}{N}\sum_{i}^{N}\|Y_{pred}^{i} - Y_{phy}^{i}\|,\\
\mathcal{L}_{phy} &= \mathcal{L}_{rc} + \mathcal{L}_{pc}
\end{align}
where $\mathcal{L}_{rc}$ denotes the reconstruction constraint part, $Y_{t}^{i}$ is the ground truths in the observable interval $[0:T]$, $\hat{Y}_{t}^{i}$ is the reconstructed results in the observable interval by the learned Richards equation. The reconstruction constraint part aims to ensure the consistency between the learned macroscopic physical laws and the observable values. $\mathcal{L}_{pc}$ denotes the prediction constraint part. This part is expressed by the mean absolute error between the neural prediction and physical prediction, which aims to lead the neural predictions closer to the physical laws. 
{The significance of the superposition of these two parts of the loss function is to enable the physical equation with learnable parameters to better approximate the real information propagation patterns. $\mathcal{L}_{rc}$ is used to make the generated values of the physical model close to the popularity of information flow within our observation interval, while $\mathcal{L}_{pc}$ is used to make the generated values of the physical model close to the predictions of the neural network.}

The clustering loss is expressed by eq.~\ref{eq:clu_loss}. The total loss function is the weighted sum of these parts, which is defined as follows:
\begin{equation}
\mathcal{L}_{total} = \alpha\mathcal{L}_{pred} + \beta\mathcal{L}_{phy} + \gamma\mathcal{L}_{clu}
\end{equation}
where $\alpha$, $\beta$ and $\gamma$ are trade-off weights of different parts in the loss function, which are all set to $1$ by default.

\section{Experiments}\label{sec:exp}
In this section, we conduct extensive experiments to demonstrate our proposed model on three public datasets to answer the following research questions:
\begin{itemize}
    \item \textbf{RQ1:} How does our proposed PIACN perform compared with other state-of-the-art baselines in information popularity prediction?
    \item \textbf{RQ2:} How does our model perform compared with different variants in the ablation study?
    \item \textbf{RQ3:} What macroscopic patterns of information cascades does our model capture through physical network modeling? How does the adaptive clustering learning mechanism capture the heterogeneity of information categories?  
    \item \textbf{RQ4:}  How do the model parameters (e.g., the hidden dimensions) affect the performance of our model?   
\end{itemize}
\subsection{Datasets and Settings}
In this paper, we evaluate our model on three public dataset: Sina Weibo, Twitter and American Physical Society (APS). The description of these three datasets are as follows:
\begin{itemize}
    \item \textbf{Sina Weibo:} This dataset was collected from the largest microblog platform in China\footnote{https://bit.ly/weibodataset}. It comprises 119,313 tweets shared on June 1, 2016, with subsequent tracking of all retweets for each post over the following 24 hours. Each tweet and its retweets create an information retweet cascade. We designated the observation time interval as 1 hour, 2 hours and 3 hours, and the prediction time as 24 hours. To mitigate the impact of users' diurnal patterns on Sina Weibo, we restrict our analysis to cascades published between 8:00 and 18:00. 
    \item \textbf{Twitter:} This dataset was collected from the largest world-wide social media platform\footnote{https://carl.cs.indiana.edu/data/\#virality2013}. It comprises publicly available English tweets written between March 24 and April 25, 2012. An information cascade is formed by a hashtag and its adopters. We specified the observation time interval as 2 days, 4 days and 6 days, and the prediction time as 32 days.
    \item \textbf{APS:} This dataset is sourced from the American Physical Society\footnote{https://journals.aps.org/datasets} and encompasses all papers published in 17 APS journals from 1893 to 2017. Each paper and its citing papers contribute to an information citation cascade. We defined the observation time interval as 3 years, 6 years and 9 years, and the prediction time as 20 years. To ensure adequate prediction intervals, we only consider papers published before 1997.
\end{itemize}

Each dataset is split with 70\% for training, 15\% for validation and 15\% for testing. Our model is implemented by Pytorch 1.5 with NVIDIA TESLA V100 GPU. 
On Weibo dataset, we set the granularity of time snapshots in observable interval as 10 minutes. On Twitter dataset, we set the granularity of time snapshots in observable interval as 8 hours. 
On APS dataset, we set the granularity of time snapshots in observable interval as 0.5 years.
The number of attention heads is set as 4, the maximum number of clusters is set as 5, and the hidden dimension of our model is set as 32.
The optimizer of our model is set as Adam~\cite{kingma2014adam}. The batch size is 64 and the learning rate is 0.0001. Our model is evaluated five times on each dataset. During training process, we utilize the early stopping strategy with tolerance 30 for 100 epochs. For each algorithm, we conduct 5 independent experiments to take the average of its metrics.

\subsection{Overall Performance (RQ1)}

We compare our model with several state-of-art baselines, including Feature-Linear~\cite{cao2017deephawkes}, XGBoost~\cite{chen2016xgboost}, MLP, DeepCas~\cite{li2017deepcas}, DeepHawkes~\cite{cao2017deephawkes} , CasCN~\cite{chen2019information} , CasFlow~\cite{xu2021casflow}, CasGCN~\cite{xu2020casgcn}, CCASGNN~\cite{wang2022ccasgnn}, I3T~\cite{tai2023predicting}, TEDDY~\cite{bao2024popularity} and POFHP~\cite{li2025public}. The descriptions of these baselines are as follows:
Similar to the work~\cite{cao2017deephawkes}, the evaluation metrics are mean square log-transformed error (MSLE) and  mean absolute percentage error (MAPE) 
averaged over five times for incremental popularity prediction. MSLE can be used to measure the overall error level of the dataset, while MAPE can measure the relative error of the dataset. 
The smaller the values of MSLE and MAPE, the better the performance of the model. 

\begin{table*}[h]
\centering
\caption{Comparison of our model with other baseline models for predicting information popularity on Weibo, Twitter and APS. The optimal metrics are highlighted in bold, while the sub-optimal metrics are underscored.}
\label{tab:comparison}
\scalebox{0.87}{\begin{tabular}{c|cc|cc|cl}
\hline
Dataset & \multicolumn{2}{c|}{Weibo (1h / 2h / 3h)} & \multicolumn{2}{c|}{Twitter (2d / 4d / 6d)} & \multicolumn{2}{c}{APS (3y / 6y / 9y)} \\ \hline
Metircs & MSLE & MAPE & MSLE & MAPE & MSLE & \multicolumn{1}{c}{MAPE} \\ \hline
Feature-Linear &3.701 / 3.365 / 3.328   & 0.529 / 0.491 / 0.457   & 2.670 / 2.139 / 1.945 & 0.433 / 0.397 / 0.361 & 3.411 / 3.116 / 2.887
 & 0.532 / 0.471 / 0.427 \\
XGBoost & 3.625 / 3.410 / 3.242 & 0.501 / 0.473 / 0.437 & 2.491 / 1.989 / 1.824 & 0.417 / 0.383 / 0.344 & 3.073 / 2.674 / 2.307 & 0.491 / 0.448 / 0.391 \\
MLP & 3.468 / 3.102 / 2.811  & 0.481 / 0.459 / 0.412 & 2.327 / 1.971 / 1.840 & 0.396 / 0.351 / 0.326 & 2.974 / 2.520 / 2.187 & 0.488 / 0.439 / 0.381 \\
DeepCas &3.631 / 3.213 / 3.107  & 0.492 / 0.461 / 0.418 & 2.019 / 1.784 / 1.539 & 0.389 / 0.337 / 0.313 & 2.719 / 2.388 / 2.018 & 0.471 / 0.424 / 0.379  \\
DeepHawkes & 2.448 / 2.279 / 2.223 & 0.468 / 0.438 / 0.382 & 1.813 / 1.507 / 1.322 & 0.358 / 0.309 / 0.291 & 2.511 / 2.174 / 1.720 & 0.442 / 0.397 / 0.334 \\
CasCN & 2.316 / 2.254 / 2.131 & 0.442 / 0.421 / 0.361 & 1.758 / 1.437 / 1.290 & 0.339 / 0.296 / 0.280 & 2.348 / 2.029 / 1.675 & 0.429 / 0.386 / 0.312 \\
CasFlow & 2.206 / 2.113 / 2.084 & 0.435 / 0.418 / 0.353 & 1.660 / 1.354 / 1.205 & 0.328 / 0.288 / 0.266 & 2.298 / 1.918 / 1.577 & 0.419 / 0.377 / 0.301\\
CasGCN & 2.374 / 2.197 / 2.108 & 0.449 / 0.421 / 0.373 & 1.714 / 1.396 / 1.281 & 0.337 / 0.291 / 0.274 & 2.288 / 2.112 / 1.659 & 0.430 / 0.389 / 0.319 \\
CCASGNN & 2.213 / 2.128 / 2.078 & 0.431 / 0.417 / 0.362 & 1.689 / 1.349 / 1.231 & 0.325 / 0.284 / 0.269 & 2.249 / 2.027 / 1.610 & 0.414 / 0.374 / 0.306 \\
I3T & \underline{2.023} / 2.156 / 1.934 & 0.436 / 0.405 / 0.353 & 1.621 / 1.376 / 1.219 & 0.323 / 0.280 / 0.265 & 2.230 / \underline{1.706} / 1.438   & 0.415 / 0.369 / 0.298 \\
TEDDY & 2.131 / 1.978 / \underline{1.820}  & 0.428 / 0.397 / 0.350 & 1.522 / \underline{1.303} / 1.197 & \underline{0.311} / \underline{0.271} / 0.260 & \underline{2.124 }/ 1.772 / \underline{1.401} & \underline{0.408} / 0.353 / 0.299 \\
POFHP & 2.078 / \underline{1.963} / 1.842 & \underline{0.425} / \underline{0.394} / \underline{0.348} & \underline{1.503} / 1.327 / \underline{1.184} & 0.312 / 0.273 / \underline{0.258} & 2.189 / 1.814 / 1.475& 0.409 / \underline{0.351} / \underline{0.298} \\
PIACN &\textbf{1.821 / 1.728 / 1.684} &\textbf{0.411 / 0.382 / 0.334}  &\textbf{1.358 / 1.148 / 1.062}  &\textbf{0.295 / 0.252 / 0.241}  & \textbf{1.917 / 1.486 / 1.232} & \textbf{0.389 / 0.335 / 0.283} \\ 
Improvement (\%) & \textbf{+9.98 / +12.64 / +7.47} & \textbf{+3.29 / +3.05 / +4.02} & \textbf{+9.64 / +11.90 / +10.30} & \textbf{+5.14 / +7.01 / +6.59} & \textbf{+9.75 / +11.97 / +12.06} & \textbf{+4.66 / +4.56 / +5.03} \\ \hline
\end{tabular}%
}
\end{table*}

The comparative experimental results are shown in Table~\ref{tab:comparison}.
From the results in these tables, we can observe that our model PIACN consistently outperforms the sub-optimal baselines with 6.45\%$\sim$12.64\% improvements in terms of MSLE and  3.05\%$\sim$7.01\% improvements in terms of MAPE on the three datasets, which demonstrates the superiority of our proposed method. In this subsection, we analyze and compare the strengths of our proposed model with other state-of-art baselines.
Among them, Feature-Linear, XGBoost, and MLP are relatively simple methods that directly input relevant features to predict incremental popularity, neglecting the learning of complex dynamics in information cascades. Therefore, their predictive performance is consistently inferior to the model we proposed.
DeepCas is the first attempt to use a recurrent neural architecture to capture the temporal dynamics of information cascades in deep learning methods. Based on DeepCas, CasCN and CasGCN further considers learning dynamic cascade structures and implements this through GCN. Although their predictive performance has improved compared to the simple methods, the lack of consideration for the physical laws governing information diffusion may be a limiting factor in their further enhancement.
DeepHawkes bridges the gap between adaptive nonlinear modeling and physical modeling through deep learning and Hawkes processes. However, since Hawkes processes only consider the microscopic dynamics of the propagation process and neglect the macroscopic patterns, they cannot accurately predict future incremental popularity.
The methods proposed in recent two years, CasFlow, I3T, TEDDY and POFHP, have improved upon previous methods by constructing more complex neural network architectures to better capture the microscopic dynamics in information cascades. However, as they focus primarily on enhancing the learning of microscopic dynamics and overlook the importance of modeling macroscopic physical laws, as well as lack attention to the heterogeneity of information categories, their predictive performance is weaker than our proposed model.

\begin{table*}[h]
\centering
\caption{Ablation Study.}
\label{tab:ablation}
\scalebox{0.9}{\begin{tabular}{c|cc|cc|cl}
\hline
Dataset & \multicolumn{2}{c|}{Weibo (1h / 2h / 3h)} & \multicolumn{2}{c|}{Twitter (2d / 4d / 6d)} & \multicolumn{2}{c}{APS (3y / 6y / 9y)} \\ \hline
Metircs & MSLE & MAPE & MSLE & MAPE & MSLE & \multicolumn{1}{c}{MAPE} \\ \hline
w/o PMN &2.075 / 1.986 / 1.827   &0.447 / 0.397 / 0.369   & 1.556 / 1.291 / 1.270 & 0.334 / 0.288 / 0.268 & 2.138 / 1.685 / 1.422
 & 0.428 / 0.364 / 0.318 \\
w/o PN & 2.117 / 2.042 / 1.930 & 0.452 / 0.409 / 0.374 & 1.652 / 1.367 / 1.298 & 0.342 / 0.293 / 0.277 & 2.245 / 1.773 / 1.561 & 0.435 / 0.372 / 0.326 \\
w/o ACN & 1.917 / 1.794 / 1.736  & 0.427 / 0.394 / 0.350 & 1.439 / 1.196 / 1.128 & 0.297 / 0.261 / 0.248 & 2.034 / 1.547 / 1.295 & 0.398 / 0.345 / 0.296 \\
w/o TLN &1.959 / 1.847 / 1.802  & 0.432 / 0.404 / 0.356 & 1.496 / 1.274 / 1.197 & 0.306 / 0.269 / 0.257 & 2.156 / 1.615 / 1.339 & 0.411 / 0.351 / 0.298  \\
w/o CEN & 1.928 / 1.814 / 1.753 & 0.426 / 0.395 / 0.347 & 1.472 / 1.239 / 1.188 & 0.298 / 0.264 / 0.251 & 2.142 / 1.603 / 1.348 & 0.402 / 0.346 / 0.293 \\
PIACN &\textbf{1.821 / 1.728 / 1.684} &\textbf{0.411 / 0.386 / 0.340}  &\textbf{1.358 / 1.148 / 1.062}  &\textbf{0.295 / 0.257 / 0.243}  & \textbf{1.917 / 1.486 / 1.232} & \textbf{0.393 / 0.338 / 0.287} \\ \hline
\end{tabular}%
}
\end{table*}


\subsection{Ablation Study (RQ2)}
To verity the effectiveness of key components in our model, we conduct ablation studies on the three datasets. As shown in Table, we compared PIACN with following variants: 1) \emph{w/o PMN}, which removes the physical modeling network from our model 2) \emph{w/o PN}, which removes the prediction network from our model  3) \emph{w/o ACN}, which removes the adaptive clustering network from our model 4) \emph{w/o TLN}, which replaces the temporal learning network with the simple MLP  5) \emph{w/o CEN}, which removes the cascade embedding network from our model.


From the experimental results, we can observe that PIACN outperforms all the ablation variants. 
Compared to \emph{w/o PMN}, PIACN significantly improves performance on Weibo, Twitter, and APS, indicating that macroscopic physical law modeling is effective for information popularity.
Compared to \emph{w/o PN}, PIACN also shows significant improvements in metrics on Weibo, Twitter, and APS. This indicates that without the neural network prediction module, it is difficult to capture the uncertainties in information popularity prediction beyond macroscopic patterns. Therefore, both the prediction network and the physical modeling network are indispensable, they complement each other and jointly improve prediction performance.
Compared to \emph{w/o ACN}, PIACN shows some improvements in performance on Weibo, Twitter, and APS, indicating that the adaptive clustering learning mechanism is effective for heterogeneous information learning.
There is also a significant fall of the performance without cascade embedding network and temporal learning network (\emph{w/o CEN} and \emph{w/o TLN}), demonstrating the effectiveness of these two parts for learning the dynamics of information cascades. 

\subsection{Case Study (RQ3)}
\begin{figure}[h]
\centering
\vspace{-2mm}
\includegraphics[width=0.48 \textwidth]{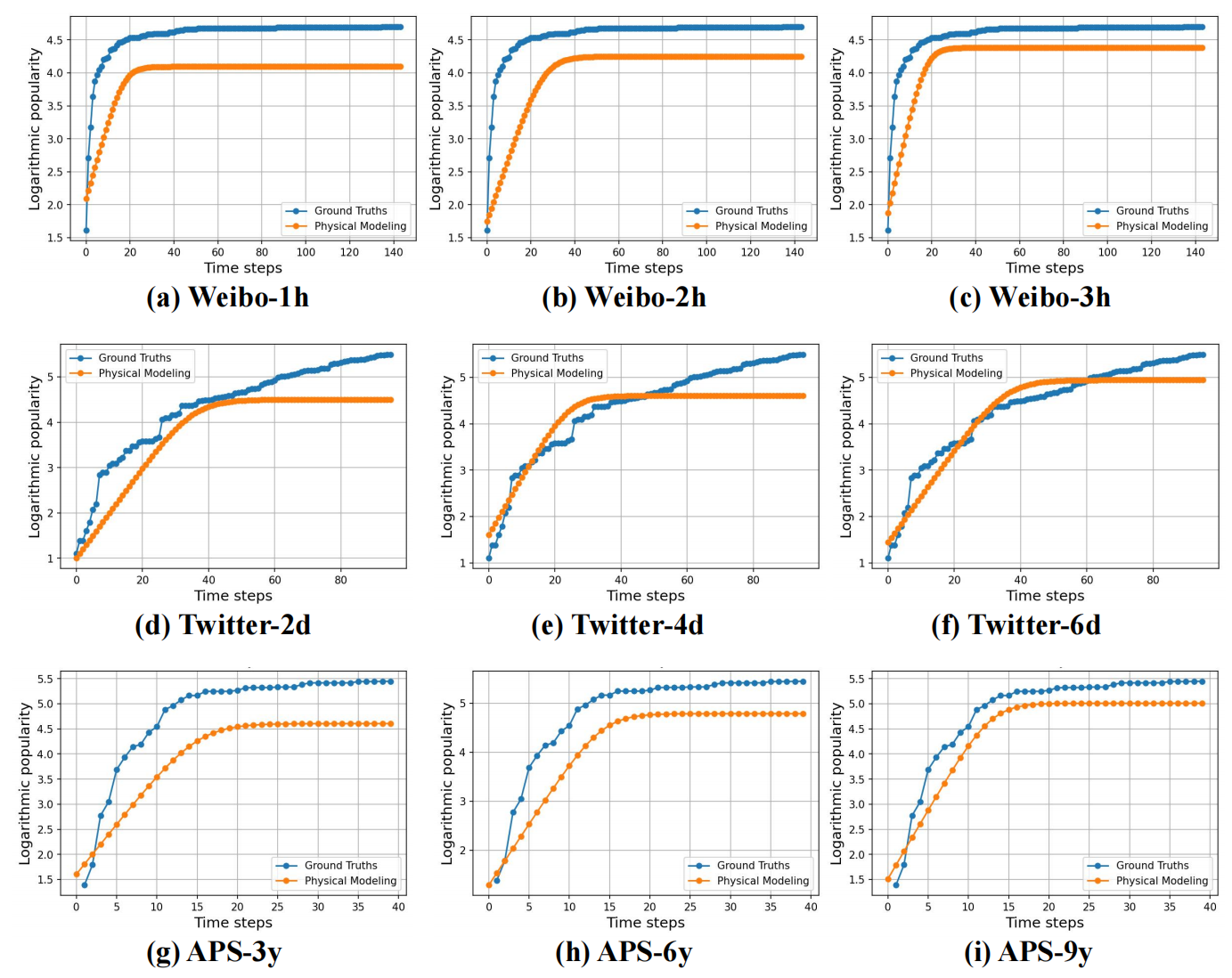}
\caption{The visualization on physical modeling curves.}
\label{fig:vis_curve} 
\vspace{-2mm}
\end{figure}

We conduct the case study on these three datasets to further investigate how the physical modeling network and adaptive clustering network work and contribute to the effectiveness of our model. As shown in Fig.~\ref{fig:vis_curve}, we compare the curve output from physical modeling network with the real curve of information popularity. In Fig.~\ref{fig:vis_curve}(a)-(c), we select a post from Weibo that was published at 10 a.m. on June 1, 2016, and visualized the output curves from the physical modeling network when the observable time interval is 1 hour, 2 hours, and 3 hours. 
In Fig.~\ref{fig:vis_curve}(d)-(f), we select a post from Twitter that was published at 10 a.m. on April 20, 2012, and visualized the output curves from the physical modeling network when the observable time interval is 2 days, 4 days, and 6 days.
In Fig.~\ref{fig:vis_curve}(e)-(i), we select an article from APS that was published in 1995, and visualized the output curves from the physical modeling network when the observable time interval is 3 years, 6 years, and 9 years.
From the visualizations on the three datasets, we can observe that as the observable time interval increases, the curves obtained from the physical modeling network become increasingly similar to the real popularity curves, especially the estimation of the final popularity becomes more and more accurate, which demonstrates the effectiveness of the physical modeling network.

\begin{figure}[h]
\centering
\vspace{-2mm}
\includegraphics[width=0.48 \textwidth]{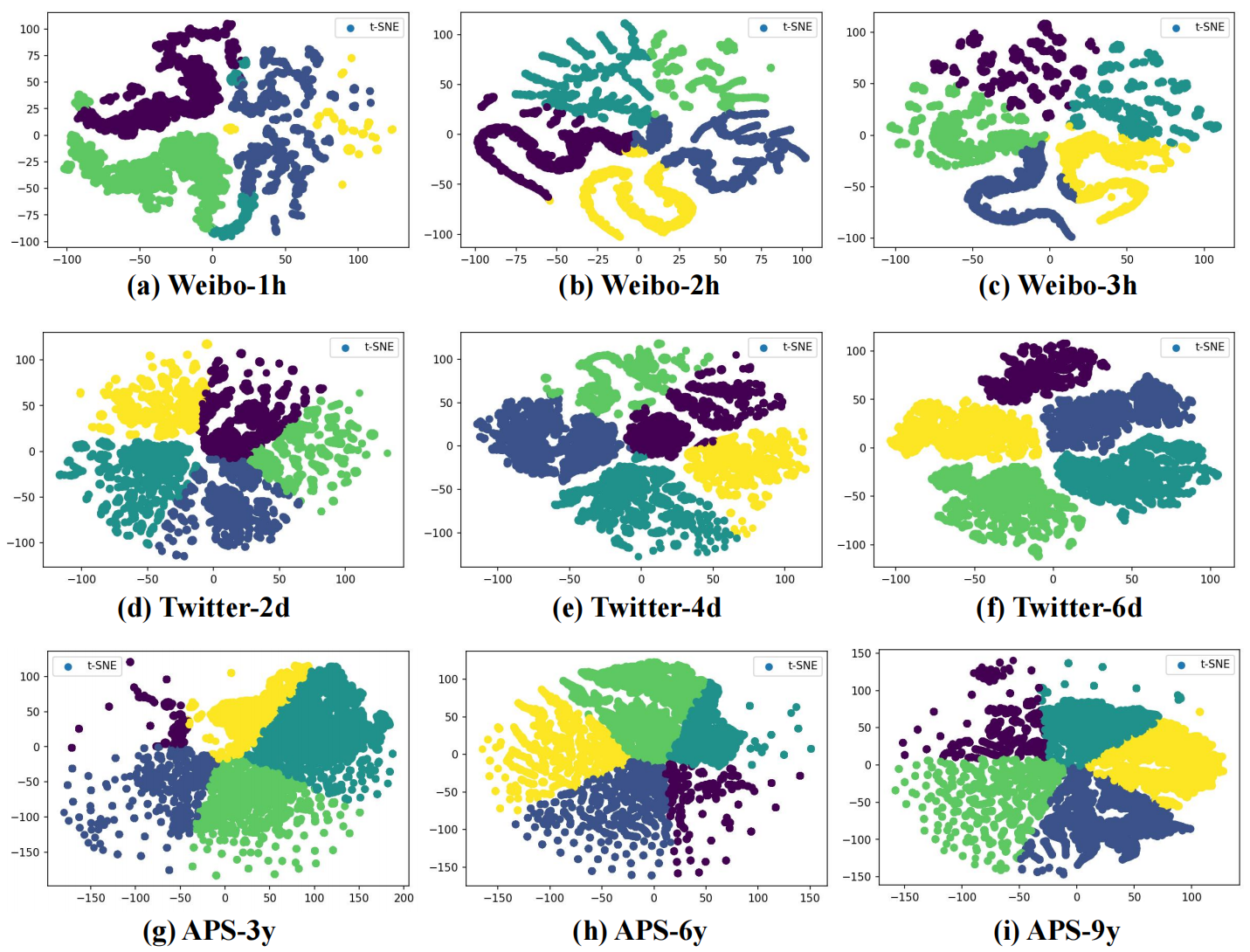}
\caption{The visualization on adaptive hidden representation clustering.}
\label{fig:vis_clu} 
\vspace{-2mm}
\end{figure}

Next, we use the T-SNE algorithm to map the high-dimensional hidden representations onto a 2D plane for visualizing the clustering clusters conveniently.
In Fig.~\ref{fig:vis_clu}, we visualize the clustering performance of the adaptive clustering network on hidden representations across different datasets. Fig.~\ref{fig:vis_clu} (a)-(c) show the clustering performance on Weibo when the observable time interval is set to 1 hour, 2 hours, and 3 hours, respectively. Fig.~\ref{fig:vis_clu} display the clustering performance on Twitter with observable time intervals set to 2 days, 4 days, and 6 days. Finally, Fig.~\ref{fig:vis_clu} demonstrate the clustering performance on APS with observable time intervals set to 3 years, 6 years, and 9 years.
We can observe that under the condition of setting the maximum cluster number to 5, the adaptive clustering network in our model consistently partitions the hidden representations into 5 clusters with clear boundaries, thereby demonstrating the effectiveness of this module.

\subsection{Parameters Study (RQ4)}
\begin{figure}[h]
\centering
\vspace{-2mm}
\includegraphics[width=0.48 \textwidth]{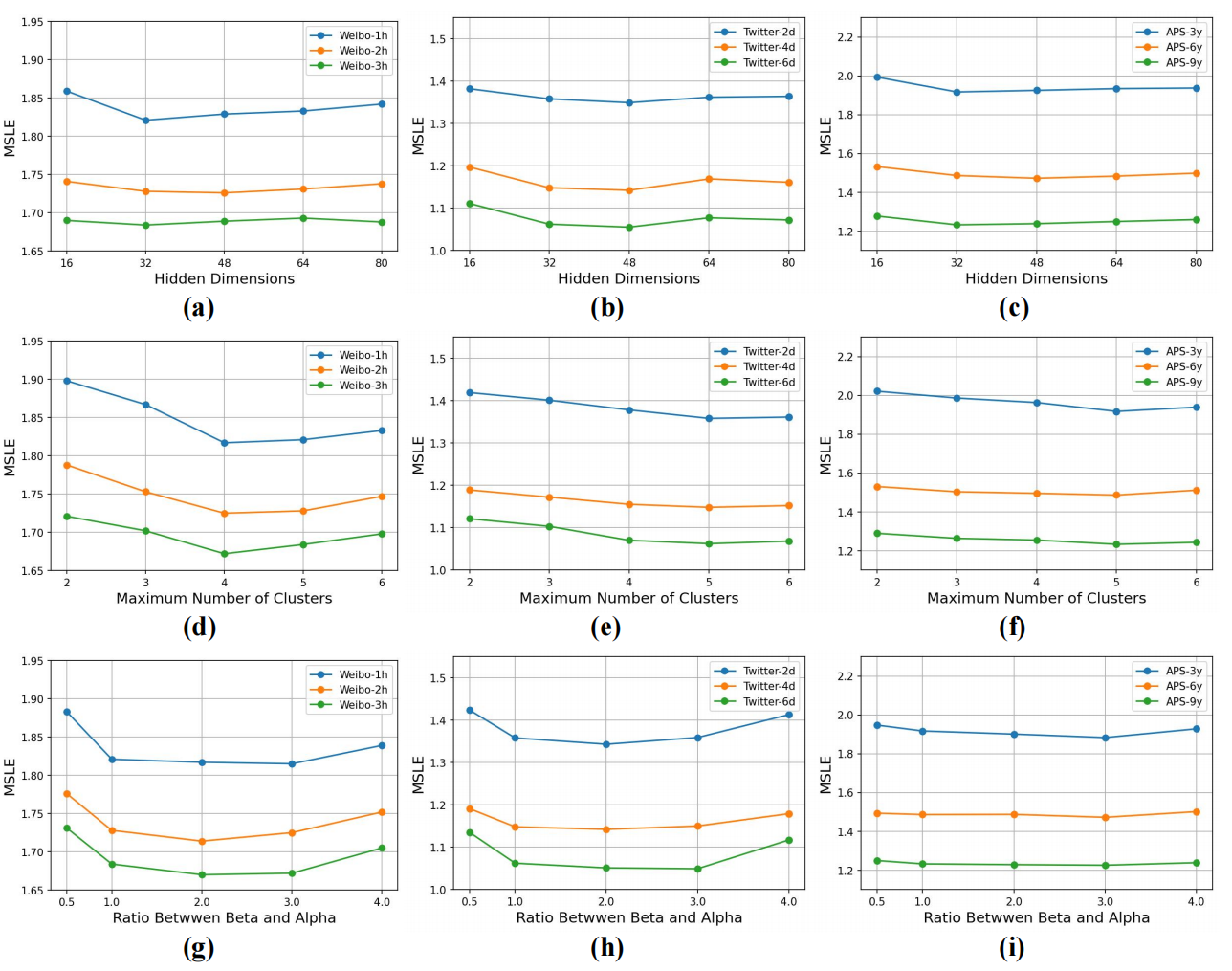}
\caption{Impact of hyper-parameters on the performance of method.}
\label{fig:param} 
\vspace{-2mm}
\end{figure}

To further investigate the effectiveness of our model, we conduct parameter study on the three datasets, including the dimension of hidden representations $D$, the maximum number of clusters $C$ and the ratio $\frac{\beta}{\alpha}$ between prediction loss and physical loss. The range of $D$ is set as$[16, 32, 48, 64, 80]$, the range of $C$ is set as $[2, 3, 4, 5, 6]$ and the range of $\frac{\beta}{\alpha}$ is set as $[0.5, 1, 2, 3, 4]$. The experimental results are shown in Fig.~\ref{fig:param}. We can find that MSLE metrics on Weibo are the optimal when $D$ is equal to 32, MSLE  metrics on Twitter are the optimal when $D$ is equal to 48, MSLE metrics on APS are the optimal when $D$ is equal to 32. If $D$ is too small, the learning capability of our model deteriorates, leading to subpar prediction performance. Conversely, when $D$ is excessively large, the metrics on the three datasets deteriorate due to over-fitting caused by an overly large hidden dimension. For parameter $C$, we can observe that MSLE metrics on Weibo achieve the optimal results when $C$ is equal to 4. On Twitter and APS, MSLE metrics obtain the best values when $C$ is equal to 5. 
This reveals that clustering the hidden representations into a moderate number of clusters can better distinguish the heterogeneity of information categories, thereby enhancing predictive performance. Given the maximum number of clusters too small or too large may both affect the learning of information category heterogeneity.
{For parameter $\frac{\beta}{\alpha}$, we fix the weight of clustering loss $\gamma$ as 1 and adjust the ratio between $\beta$ and $\alpha$. We can observe that MSLE metrics on Weibo, Twitter and APS achieve the optimal results when $\frac{\beta}{\alpha}$ is equal to 2 or 3. This reveals that appropriately increasing the weight of physical loss during training may enable the model to achieve better prediction performance.}

\section{conclusion}\label{sec:conclusion}
We propose a novel physics-informed neural network with adaptive clustering learning mechanism for information popularity prediction.
Our model not only involve the macroscopic physical laws to guide the training process by physical modeling network, but also consider the impact of information category heterogeneity through adaptive clustering network.
Extensive experiments on three real-world datasets demonstrate the superiority of our proposed model in prediction accuracy compared with other state-of-art baselines. 
In this paper, we give the first attempt to integrate macroscopic physical laws into deep learning models for predicting information popularity. 
However, our work still has two main limitations. Firstly, the Richards growth equation can only describe the general pattern of information popularity, neglecting information that may exhibit multi-modal characteristics, such as public sentiment triggered by major events. Secondly, the datasets used in our experiments does not contain the true category labels of the information, thus limiting the interpretability of the adaptive clustering learning mechanism. 
In the future, we will further optimize the physical modeling and deep model architecture, and mine more data rich in label information, so that the deep learning model can not only cope with more special cases of information propagation but also possess stronger interpretability.

\ifCLASSOPTIONcaptionsoff
  \newpage
\fi

\bibliographystyle{IEEEtran}
\bibliography{IEEEabrv,Bibliography}

\begin{IEEEbiography}
[{\includegraphics[width=1.0in,height=1.5in,clip,keepaspectratio]{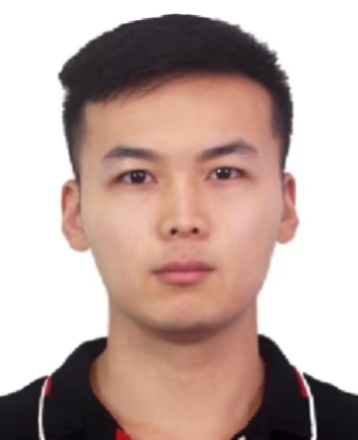}}]{Guangyin Jin} is an Assistant Research Fellow in National Innovative Institute of Defense Technology, and an IEEE Senior Member. He received a Ph.D degree at College of Systems Engineering of National University of Defense Technology. 
His research interest falls in the area of spatial-temporal data mining, graph neural networks and urban computing. So far, he has published more than 40 papers in top journals and conferences such as TKDE, TITS, TIST, TRC, TKDD, AAAI, CIKM, NIPS, ICCV, SIGSPATIL, accumulating more than 3,000 citations on Google Scholar. He is also an editorial board member for SCI-indexed journals such as Scientific Reports, Humanities \& Social Sciences Communications, and Mathematics.
\end{IEEEbiography}

\begin{IEEEbiography}
[{\includegraphics[width=1.0in,height=1.5in,clip,keepaspectratio]{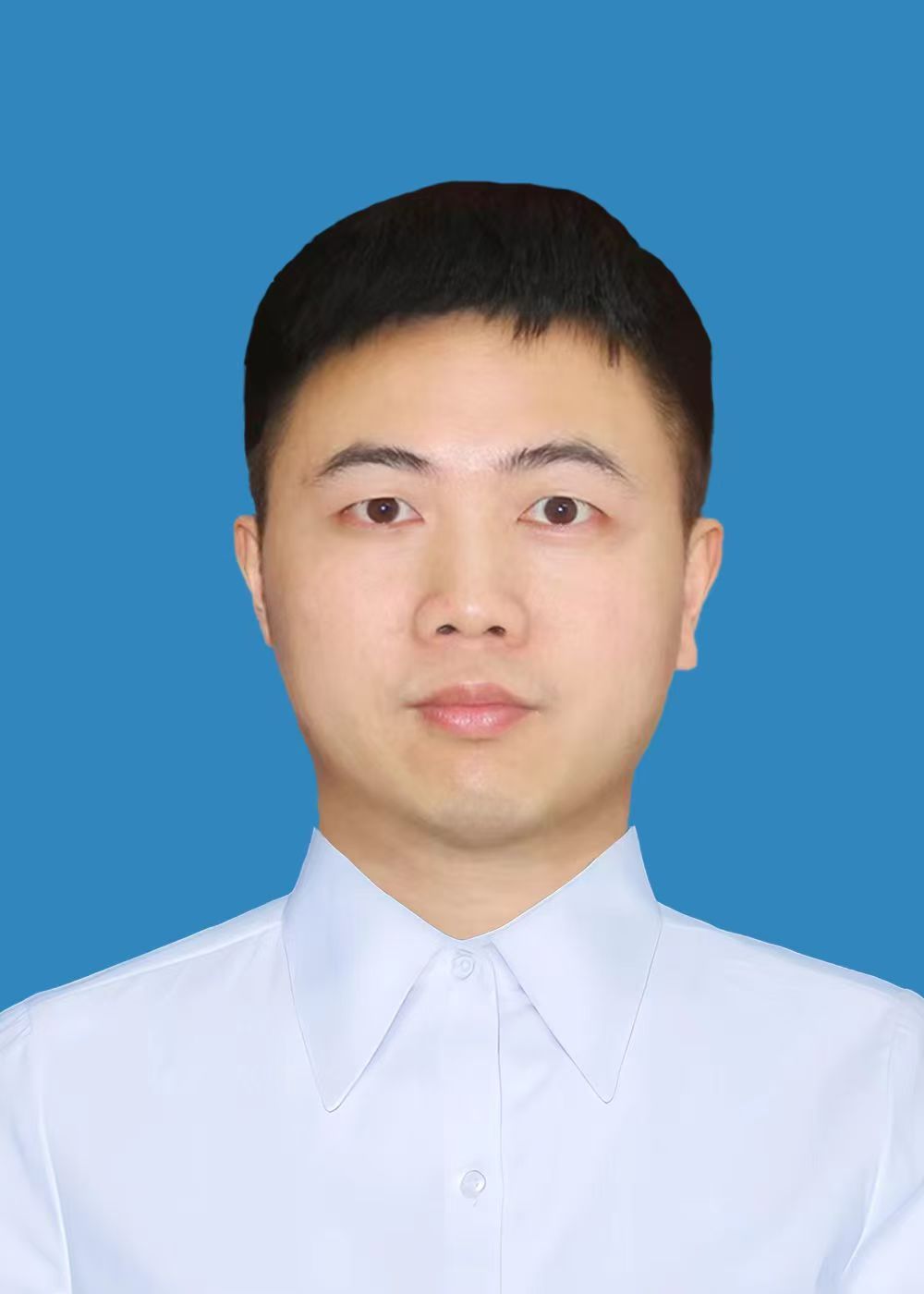}}]{Xiaohan Ni} is a postdoctoral fellow at the Capital Normal University, a lecturer at the Chinese  Armed Police Force Command College, and holds a doctoral degree from the Academy of Military Sciences. His main research interests include resources-targets matching optimization and  command analysis of operations. So far, he has published several papers in JCR Q1-level international journals such as MAA.
\end{IEEEbiography}

\begin{IEEEbiography}
[{\includegraphics[width=1.0in,height=1.5in,clip,keepaspectratio]{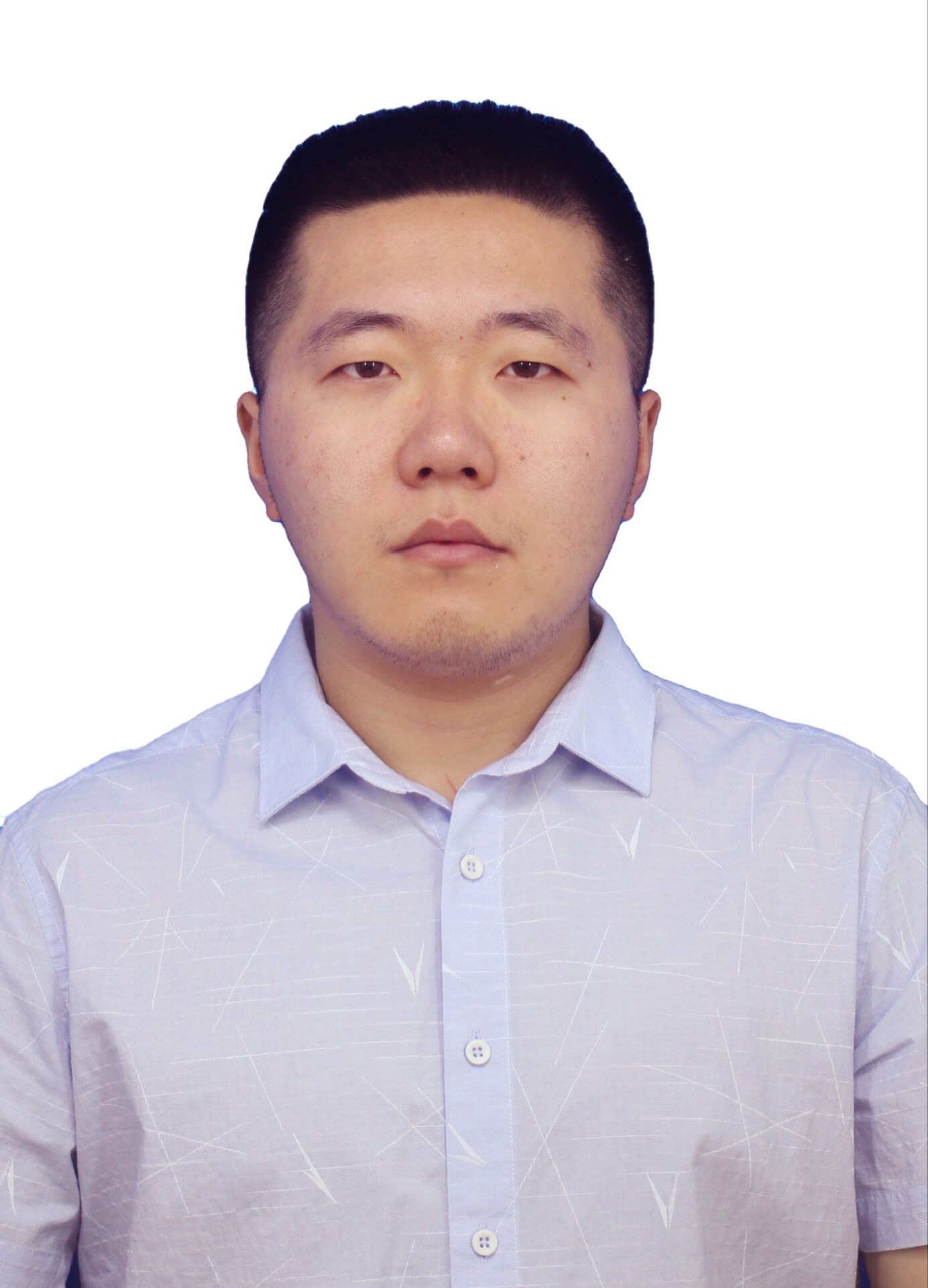}}]{Yanjie Song} received double B.S. degrees in Management from Tianjin University, Tianjin, China, in 2017 and the Ph.D. degree in Engineering from National University of Defense Technology, China, in 2023. He has published more than 60 papers in IEEE TEVC, IEEE TGRS, and other journals. He has authored 4 academic book, obtained 10 National Invention Patents, and hosted/participated in more than 15 projects. His research interests include computational intelligence, evolutionary algorithm, combinatorial optimization, and deep reinforcement learning. He is now the Guest Editor of the Swarm and Evolutionary Computation, Guest Editor of the Computers \& Electrical Engineering, Associate Editor of the International Journal on Interactive Design and Manufacturing, Young Editorial Board Member of Data Science and Management, reviewer of the IEEE TEVC, IEE TCYB, and more than 30 other journals. 
\end{IEEEbiography}

\begin{IEEEbiography}
[{\includegraphics[width=1.0in,height=1.5in,clip,keepaspectratio]{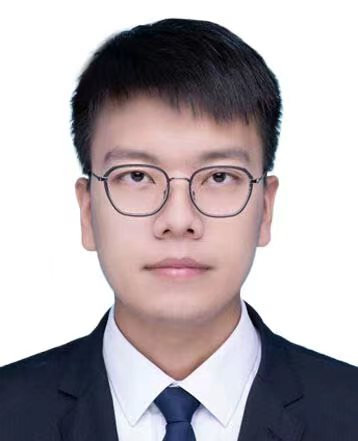}}]{Kun Wei} graduated from the School of Physics, Peking University in 2015 with a bachelor's degree, and obtained his doctorate from the Department of Engineering Physics at Tsinghua University in 2021. During his doctoral studies, he mainly engaged in research related to the interaction between radiation and matter and the measurement and diagnosis of radiation. Now, he is engaged in research on complex systems, especially social physics, aiming to achieve the representation of individual and group behaviors in society through physical modeling and computational simulation, and to explain and analyze related social phenomena to serve the strategic needs of national security and scientific governance. 
\end{IEEEbiography}

\begin{IEEEbiography}
[{\includegraphics[width=1.0in,height=1.5in,clip,keepaspectratio]{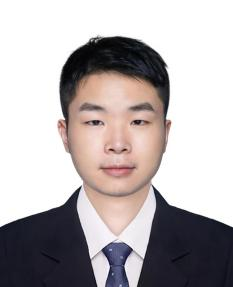}}]{Jie Zhao} is a PhD in Physics with a robust interdisciplinary background in plasma physics and optics. His research encompasses the transport of carriers in topological insulators, gamma-vortex generation, and the production and acceleration of positrons. Jie possesses extensive expertise in large-scale numerical simulations, ultrafast spectroscopy characterization, Raman spectroscopy measurements, and fluorescence upconversion experiments. Recently, he proposed a novel physical scheme for all-optical positron production and acceleration, which has garnered widespread recognition in the field and was published in the prestigious journal Communication Physics. Currently, Jie serves as an assistant researcher in the field of complex systems and social physics, where he investigates the intricate interactions and dynamics that govern these systems. 
\end{IEEEbiography}

\begin{IEEEbiography}
[{\includegraphics[width=1.0in,height=1.5in,clip,keepaspectratio]{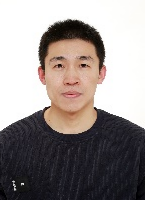}}]{Leiming Jia} is a Ph.D. in Nuclear Science and Technology with a solid interdisciplinary foundation, graduated from the Department of Engineering Physics of Tsinghua University. Jia is mainly engaged in the field of complexity science and nonlinear dynamics, and has constructed theoretical calculation methods for shock wave propagation in complex environments. Recently, Jia mainly conducts researches on three-party (multi-party) game theory, including the research and establishment of physical models and computational models, to understand the dynamic behaviors and physical mechanisms behind complex games in the real world, and the important influence of complex factors such as information and cognition. 
\end{IEEEbiography}

\begin{IEEEbiography}
[{\includegraphics[width=1.0in,height=1.5in,clip,keepaspectratio]{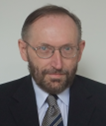}}]{Witold Pedrycz} (IEEE Life Fellow) is Professor in the Department of Electrical and Computer Engineering, University of Alberta, Edmonton, Canada. He is also with the Systems Research Institute of the Polish Academy of Sciences, Warsaw, Poland. Dr. Pedrycz is a foreign member of the Polish Academy of Sciences and a Fellow of the Royal Society of Canada. He is a recipient of  several awards including Norbert Wiener award from the IEEE Systems, Man, and Cybernetics Society, IEEE Canada Computer Engineering Medal, a Cajastur Prize for Soft Computing from the European Centre for Soft Computing, a Killam Prize, a Fuzzy Pioneer Award from the IEEE Computational Intelligence Society, and 2019 Meritorious Service Award from the IEEE Systems Man and Cybernetics Society. His main research directions involve Computational Intelligence, Granular Computing, and Machine Learning, among others. Professor Pedrycz served as an Editor-in-Chief of Information Sciences, Editor-in-Chief of WIREs Data Mining and Knowledge Discovery (Wiley), and Co-editor-in-Chief of Int. J. of Granular Computing (Springer) and J. of Data Information and Management (Springer). Currently he serves as Editor-in-Chief of WIREs Data Mining and Knowledge Discovery (Wiley). 
\end{IEEEbiography}

\vfill


\end{document}